\documentclass[showpacs,12pt,preprintnumbers,amsmath,amssymb,floatfix]{revtex4}
\usepackage{graphicx}
\usepackage{dcolumn}
\usepackage{bm}
\def\be{{\bf e}}
\def\bF{{\bf F}}
\def\bk{{\bf k}}

\def\bnh{{\bf \hat{n}}}
\def\br{{\bf r}}

\def\brh{{\bf \hat{r}}}

\def\bu{{\bf u}}

\def\bV{{\bf V}}
\def\bW{{\bf W}}
\def\bx{{\bf x}}
\def\bX{{\bf X}}
\def\cH{{\cal H}}
\def\cL{{\cal L}}

\def\hx{{\hat x}}
\def\hy{{\hat y}}
\def\hz{{\hat z}}
\def\eikr{e^{i\bk\cdot\br}}

\def\ylm{{Y_{lm}}}

\begin{document}

\title{Soft modes near the buckling transition of icosahedral shells}

\author{M. Widom}
\affiliation{
Department of Physics, Carnegie Mellon University\\
Pittsburgh, PA 15213  USA\\
Department of Computational Biology, School of Medicine\\
University of Pittsburgh,
Pittsburgh, PA  15213  USA}
\author{J. Lidmar}
\affiliation{Department of Physics, Royal Institute of Technology\\
AlbaNova, SE-106 91 Stockholm  Sweden}
\author{David R. Nelson}
\affiliation{Department of Physics, Harvard University\\
Cambridge, MA 02139  USA}
\date{\today}

\begin{abstract}
Icosahedral shells undergo a buckling transition as the ratio of
Young's modulus to bending stiffness increases.  Strong bending
stiffness favors smooth, nearly spherical shapes, while weak bending
stiffness leads to a sharply faceted icosahedral shape.  Based on the
phonon spectrum of a simplified mass-and-spring model of the shell, we
interpret the transition from smooth to faceted as a soft-mode
transition.  In contrast to the case of a disclinated planar network
where the transition is sharply defined, the mean curvature of the
sphere smooths the transitition.  We define elastic susceptibilities
as the response to forces applied at vertices, edges and faces of an
icosahedron.  At the soft-mode transition the vertex susceptibility is
the largest, but as the shell becomes more faceted the edge and face
susceptibilities greatly exceed the vertex susceptibility.  Limiting
behaviors of the susceptibilities are analyzed and related to the
ridge-scaling behavior of elastic sheets.  Our results apply to virus
capsids, liposomes with crystalline order and other shell-like
structures with icosahedral symmetry.
\end{abstract}

\maketitle

\section{Introduction}

\begin{figure}
\includegraphics[width=3in,angle=180]{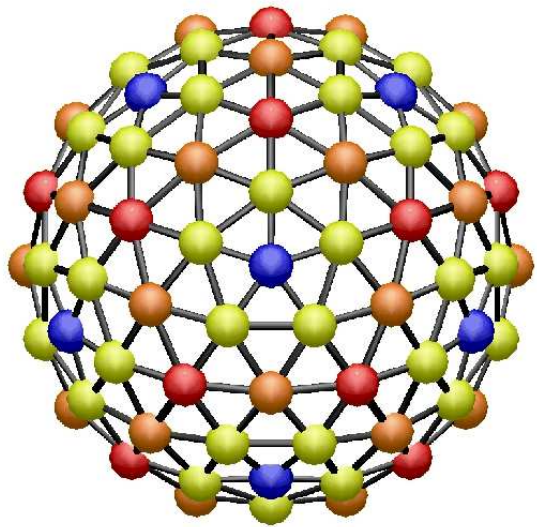}
\caption{Triangulated network of $P=Q=2$.  Colors identify local environments
with 5-fold vertices shown in blue.}
\label{fig:mass+spring}
\end{figure}

Virus capsids~\cite{Caspar62} and other structures such as
colloidosomes~\cite{Dinsmore} and liposomes~\cite{Spector,Delorme}
consist of thin shells of spherical topology that frequently exhibit
icosahedral symmetry.  A popular simplified
model~\cite{Lidmar03,Toan06,Vlieg06,Henley06} replaces the shell with
a triangulated network of masses and springs (see
Fig.~\ref{fig:mass+spring}).  This network consists of five- and
six-coordinated vertices, with the five-coordinated vertices aligned
with the five-fold icosahedral symmetry axes.  Five-coordinated
vertices may be considered as $+2\pi/6$ disclinations within an
otherwise six-coordinated lattice.  These disclinations are absent in
conventional continuum models of spherical
shells~\cite{Timo,Love,Niordson}.

\begin{figure}
\includegraphics[width=3in,angle=180]{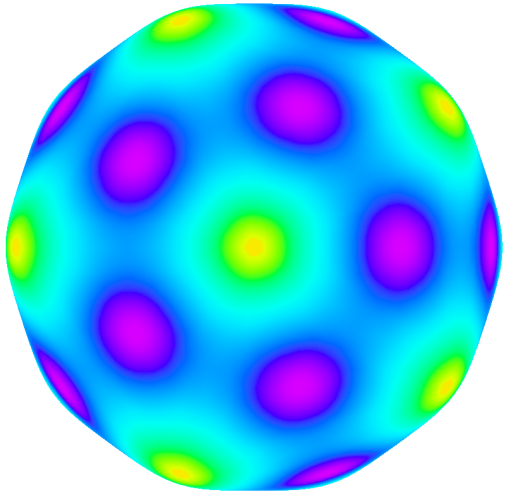}
\caption{Shell shape above the buckling transition for $P=128$, $Q=0$
shell with $k_s=1$ and $k_b=16$ yielding Foppl-von Karman number
$\gamma$=930.  Color coding is logarithmic according to total elastic
energy (violet=low, red=high)}.
\label{fig:shells}
\end{figure}

Elastic properties of the capsid can be mimicked by suitably adjusting
the spring constants to obtain the desired Young's modulus $Y$ and by
imposing a curvature energy to obtain the bending modulus $\kappa$.
Strains associated with the disclinations cause the network to
buckle~\cite{Seung88}, transforming the shape from smooth and nearly
spherical to strongly faceted and nearly icosahedral~\cite{Lidmar03}.
A dimensionless parameter controls the transformation.  We define the
Foppl-von Karman number
\begin{equation}
\label{eq:FvK}
\gamma=\frac{Y R^2}{\kappa}
\end{equation}
where $R$ is a linear dimension of the shell.  The buckling occurs
when $\gamma$ exceeds a value $\gamma_b$ of order $10^2$ (see
Fig.~\ref{fig:shells}).  For the virus HK97, which appears to facet as
it matures~\cite{Hendrix,Schmidt04}, $\gamma$ reaches a value of order
$10^3$ according to the estimate of Ref.~\cite{Lidmar03}.  Varying the
pH of solution can alter $\gamma$, with the range 100-900 reported for
the virus CCMV~\cite{Klug,Tama02}.  At much larger values of $\gamma$
(in excess of $10^6$) which should characterize liposomes with
crystalline order, an interesting phenomenon known as ``ridge
scaling'' emerges~\cite{Wit93,Lob95,DiD01,DiD02,Wood02,Wit07}.

Caspar and Klug~\cite{Caspar62} classify icosahedral structures by a
pair of integers $(P,Q)$.  A pair of five-coordinated vertices is
connected by a path consisting of $P$ edges in some given direction
and $Q$ edges in a direction $60^{\circ}$ to the left (e.g. between
two blue vertices {\em via} a red vertex in
Fig.~\ref{fig:mass+spring}).  The $T$-number of the network,
$T=P^2+PQ+Q^2$ gives the number of vertices as $N_v=10T+2$.  There
are always 12 five-coordinated vertices, so the number of
six-coordinated vertices is $10(T-1)$.  Structures with $P$ and $Q$
both nonzero and $P\ne Q$ are chiral, such that $(P,Q)$ and $(Q,P)$ are
mirror images.  Their symmetry group is the 60-element icosahedral
rotation group $Y$.  Structures with either $P$ or $Q=0$, or with
$P=Q$ are nonchiral.  Their symmetry belongs to the 120-element group
$Y_h=Y\times Z_2$, which should not be confused with the 120-element
icosahedral double group~\cite{Wid86} $Y'$.

We exploit the rotational symmetry group to analyze the normal modes
of the network model by diagonalizing the Hessian matrix of the
elastic energy.  Eigenvectors represent characteristic modes of
deformation, which transform according to irreducible representations
of $Y$, and the corresponding eigenvalues measure the mechanical
stability.  Because the buckling occurs in a symmetric fashion, the
corresponding modes must exhibit full icosahedral symmetry.
Nondegenerate modes transform as the unit representation.  Tracking
these nondegenerate eigenvalues reveals a softening and also a mixing
of modes as $\gamma$ passes through $\gamma_b$.

Other studies consider more microscopis elastic network
models~\cite{Bahar,Tama} that place nodes at every $C^{\alpha}$ atom
in the amino acid chains.  These studies find that the displacements
during maturation (i.e. as the virus goes through the buckling
transition) can be accurately represented using a superposition of
only the lowest few nondegenerate modes, consistent with our
expectations.

Section~\ref{sec:continuum} of this paper reviews the
continuum-elastic theory for deformations of planes and spheres, to
establish notation and for comparison with our later numerical
results.  Our network model is defined in section~\ref{sec:discrete}
and applied to the special cases of disclination-free triangular
lattices, single disclinations of positive charge, and icosahedral
structures of spherical topology containing twelve disclinations.
Low-lying eigenvalue spectra reveal a sharp buckling transition in the
case of a single disclination but a broadly smeared transition for the
icosahedral case.  Following Ref.~\cite{Lidmar03} we find that the
positive curvature of the sphere plays a symmetry-breaking role
analagous to an applied magnetic field at a ferromagnetic phase
transition.

The final section~(\ref{sec:response}) applies forces to selected
points on a plane or a shell to probe the elastic response of the
network as a whole.  The resulting displacements define
susceptibilities which diverge in the case of the single disclination.
In the case of the icosahedron, we find the effective stiffness
(inverse of the susceptibility) drops most rapidly at $\gamma_b$ for
forces applied at five-fold symmetry axes, but the stiffness falls off
more rapidly for forces applied at two- and three-fold symmetry axes
for $\gamma>\gamma_b$.  We analyze these susceptibilities in limiting
cases of small and large $\gamma$.

\section{Continuum-elastic theory}
\label{sec:continuum}

The general elasticity theory of membranes can be expressed in
coordinate-free form~\cite{Peterson84,Niordson}.  Let $M$ be a
manifold (a two-dimensional smooth surface embedded in three
dimensional space) assumed to be in mechanical equilibrium.  Now
impose tangential deformation $\bu(\bx)$ and normal deformation
$\zeta(\bx)$ corresponding to displacements of points $\bx$ on the
surface.  Let $g_{\alpha\beta}$ and $C_{\alpha\beta}$ be the metric
and curvature tensors respectively of $M$ after distortion.  Greek
indices take values 1 and 2 corresponding to the dimensions of $M$.
Define the strain tensor
\begin{equation}
U_{\alpha\beta}=u_{\alpha\beta}+\zeta C_{\alpha\beta}
\end{equation}
where $u_{\alpha\beta}=\frac{1}{2}(D_{\alpha} u_{\beta}+D_{\beta} u_{\alpha})$
and $D_{\alpha}$ indicates covariant differentiation with respect to
$x_{\alpha}$.
The trace $U_{\gamma}^{\gamma}$ measures dilation, while
\begin{equation}
S_{\alpha\beta}=U_{\alpha\beta}-\frac{1}{2}g_{\alpha\beta}U_{\gamma}^{\gamma}
\end{equation}
measures shear strain.  Bending of $M$ is characterized by mean
curvature $H=\frac{1}{2}C_{\gamma}^{\gamma}$ and Gaussian curvature
$K=\det{C}$.

The free energy density at $\bx$ contains dilation, shear and bending
contributions,
\begin{equation}
f(\bx)=f_d+f_s+f_b
\end{equation}
and can be integrated over $M$ to obtain the total free energy
\begin{equation}
F=\int f(\bx) \sqrt{\det{g}} d^2\bx.
\end{equation}
The separate contributions are
\begin{eqnarray}
f_d(\bx)=\frac{1}{2}(\lambda+\mu) (U_{\gamma}^{\gamma})^2\\ \nonumber
f_s(\bx)=\mu S^{\alpha\beta}S_{\alpha\beta}\\
f_b(\bx)=\frac{1}{2}\kappa(2H-c_0)^2+\kappa_G K. \nonumber
\end{eqnarray}
The elastic constants $\lambda$ and $\mu$ are the Lame
constants~\cite{LandL}.  The 2D area (bulk) modulus $B=\lambda+\mu$,
while $\mu$ itself is the shear modulus, and the 2D Young's modulus
$Y=4\mu(\lambda+\mu)/(\lambda+2\mu)$.  Upon integration over the
surface $M$, the Gaussian curvature term becomes constant, and we
neglect this term henceforth.  Likewise, we set the spontaneous
curvature $c_0=0$, thus assuming the manifold $M$ would be flat in the
absence of constraints associated with the spherical topology.
Effects of $c_0\ne 0$ are discussed in Ref.~\cite{Toan06}.
 
Given the elastic free energy $F$ we obtain the stress tensor
\begin{equation}
\sigma^{\alpha\beta}(\bx)=\frac{\delta F}{\delta U_{\alpha\beta}(\bx)}
\end{equation}
whose divergence yields the tangential force
\begin{equation}
F^\beta=D_\alpha\sigma^{\alpha\beta}.
\end{equation}
The normal force is given by
\begin{equation}
N(\bx)=-\frac{\delta F}{\delta \zeta(\bx)}.
\end{equation}
In mechanical equilibrium the stress tensor and normal force vanish.
Slightly out of equilibrium, to first order in the displacements,
special forms of $\bu(\bx)$ and $\zeta(\bx)$ known as normal modes
solve the eigenvalue equation
\begin{equation}
\label{eq:eig}
(\bF,N)=-\Lambda(\bu,\zeta).
\end{equation}
When displaced from equilibrium according to the $k^{th}$  normal mode
$(\bu_k,\zeta_k)$, the free energy increases by
\begin{equation}
\Delta F = \frac{1}{2}\Lambda_k\int(|\bu_k|^2+\zeta_j^2)d\bx.
\end{equation}
According to the equipartition theorem the modes fluctuate with
thermal energy $\Delta F=k_B T/2$ and amplitude
$\int(|\bu_k|^2+\zeta_k^2)d\bx=2k_B T/\Lambda_k$.

Time dependence of the strains depends on the equations of motion.  In
the overdamped case we write
\begin{eqnarray}
\dot{u}^\beta(\bx) = \Gamma F^{\beta}(\bx),
&
\dot{\zeta}(\bx) = \Gamma N(\bx)
\end{eqnarray}
where we take $\Gamma$ proportional to an inverse viscosity as in a
Stokes-Einstein relation.  In this case a normal mode decays in time
with a decay rate $\omega=\Gamma\Lambda$.  Ref~\cite{Levine02} carries
out a more thorough investigation of flat membranes coupled to fluid
flow.  In the absence of damping we write
\begin{eqnarray}
\rho \ddot{u}^\beta(\bx) = F^\beta(\bx),
&
\rho \ddot{\zeta}(\bx) = N(\bx)
\end{eqnarray}
with $\rho$ the 2D mass density.  A normal mode now oscillates in time at
frequency $\omega=\sqrt{\Lambda/\rho}$.

\subsection{Deformations of a Plane}
\label{sec:ContPlane}

An infinite flat elastic sheet in equilibrium has no curvature, so for
small perturbations the energy decouples into contributions from the
in-plane strain $\bu$ and perpendicular displacement $\zeta$.
\begin{equation}
f=\frac{1}{2}\lambda (u_\gamma^\gamma)^2
+\mu u^{\alpha\beta}u_{\alpha\beta}
+\frac{1}{2}\kappa (\Delta\zeta)^2
\end{equation}
Here $\Delta=D_\alpha D^\alpha=\nabla^2$ is the usual 2D laplacian
operator and $\nabla$ the usual gradient.  By differentiating the
energy we obtain the forces
\begin{equation}
\label{FlatIn}
\bF = (\lambda+\mu) \nabla \nabla\cdot\bu+
\mu\Delta\bu
\end{equation}
and
\begin{equation}
\label{FlatOut}
N = -\kappa\Delta^2 \zeta
\end{equation}

Because the in-plane and out-of-plane displacements decouple, we solve them
separately.  The solutions are based on the plane wave function
\begin{equation}
\label{planewave}
\psi_{\bk}(\br)=\eikr
\end{equation}
which is an eigenfunction of the Laplacian operator $\Delta
\psi_{\bk}(\br)=-k^2\psi_{\bk}(\br)$.  In-plane normal modes are
expressed as longitudinal waves
\begin{equation}
\label{flatUL}
\bu_L(\br)=\nabla \psi_{\bk}(\br)=i\bk\eikr
\end{equation}
and transverse waves
\begin{equation}
\label{flatUT}
\bu_T(\br)=\hz\times\bu_L=i(k_x\hy-k_y\hx)\eikr .
\end{equation}
Note the identities $\nabla\times\bu_L=0$ and $\nabla\cdot\bu_T=0$, as
expected for longitudinal and transverse waves.  These waves are
eigenvectors of the in-plane force eq.~(\ref{FlatIn}) provided their
eigenvalues obey the longitudinal and transverse dispersion relations,
respectively
\begin{equation}
\label{flatWL}
\Lambda_L=(\lambda+2\mu) k^2
\end{equation}
and
\begin{equation}
\label{flatWT}
\Lambda_T=\mu k^2.
\end{equation}
Perpendicular out-of-plane waves
\begin{equation}
\label{flatUP}
\bu_P(\br)=\hz\psi_{\bk}(\br)
\end{equation}
obey eq.~(\ref{FlatOut}) subject to the
perpendicular wave dispersion relation
\begin{equation}
\label{flatWP}
\Lambda_P = \kappa k^4.
\end{equation}

For future reference we recast the normal modes in plane-polar
coordinates $(r,\phi)$, replacing the plane-wave function
$\psi_{\bk}(\br)$ with cylindrical Bessel functions
\begin{equation}
\label{Jmkr}
\psi_{km}(r,\phi)=J_m(kr)e^{im\phi} .
\end{equation}
The Laplacian operator takes the form
\begin{equation}
\Delta=
\frac{1}{r}\frac{\partial}{\partial r}
\left(r\frac{\partial}{\partial r}\right)
+\frac{1}{r^2}\frac{\partial^2}{\partial \phi^2} .
\end{equation}
Like the plane wave function $\psi_{\bk}(\br)$, waves of
type~(\ref{Jmkr}) are eigenfunctions of the Laplacian operator,
$\Delta \psi_{km}(r,\phi)=-k^2\psi_{km}(r,\phi)$.  Upon defining
normal modes $\bu_L,\bu_T,\bu_P$ as in eqs.~(\ref{flatUL}),
(\ref{flatUT}) and (\ref{flatUP}) the longitudinal, transverse and
perpendicular dispersion relations given in eq.~(\ref{flatWL}),
(\ref{flatWT}) and (\ref{flatWP}) result.  These polar forms
generalize nicely to conical and spherical geometries.

\subsection{Deformations of a Sphere}
\label{sec:Cont:Sphere}

Now we redo the prior calculation of section~\ref{sec:ContPlane} for
the case of small perturbations around a sphere of equilibrium radius
$R$.  In this case the unperturbed manifold has constant mean
curvature $H_0=1/R$.  Consequently the free energy acquires a term
coupling the in-plane and normal strains through the dilation energy.
\begin{eqnarray}
f_d=\frac{1}{2}(\lambda+\mu)(u_\gamma^\gamma+2\zeta/R)^2\\ \nonumber
f_s=\mu(u_{\alpha\beta}u^{\alpha\beta}-(u_\gamma^\gamma)^2)\\
f_b\sqrt{\det{g}}=\frac{1}{2}\kappa\left(
(\frac{2}{R}-\Delta \zeta)^2
-\frac{2}{R^2}(D_\alpha \zeta)^2
\right) \nonumber
\end{eqnarray}
In the above, the Laplacian operator takes the form
\begin{equation}
\Delta=\frac{1}{R^2 \sin{\theta}}
\frac{\partial}{\partial\theta}
\left(\sin{\theta}\frac{\partial}{\partial\theta}\right)
+\frac{1}{R^2\sin^2{\theta}}\frac{\partial^2}{\partial\phi^2}.
\end{equation}
Notice we include the integration measure $\sqrt{\det{g}}$ along with
the bending energy $f_b$, because it contributes the term
$(D_\alpha\zeta)^2$.  The $\sqrt{\det{g}}$ factor is not needed in $f_d$
or $f_s$ because these are already second order in the deformation.

Taking functional derivatives of $F$ yields the stress tensor, in-plane
and normal force
\begin{eqnarray}
\label{forces}
\sigma^{\alpha\beta}=
\lambda g^{\alpha\beta}(u_\gamma^\gamma+\frac{2}{R}\zeta)
+2\mu(u^{\alpha\beta}+\frac{1}{R}g^{\alpha\beta}\zeta)\\ \nonumber
F^\beta=(\lambda+\mu)D^\beta(u_\gamma^\gamma+\frac{2}{R}\zeta)
+\mu(\Delta+\frac{1}{R^2}) u^\beta\\
N=-(\lambda+\mu)(\frac{2}{R}u_\gamma^\gamma+\frac{4}{R^2}\zeta)
-\kappa {\cL} \zeta \nonumber
\end{eqnarray}
where we define
\begin{equation}
\label{cL}
\cL=D_\alpha D^\alpha D_\beta D^\beta+\frac{2}{R^2}D_\alpha D^\alpha .
\end{equation}
The extra $\mu u^\beta/R^2$ in eq.~(\ref{forces}) for $F^{\beta}$ comes
from commutation of covariant derivatives.  The final, second
derivative, term in~(\ref{cL}) comes from integrating by parts the
square of the first derivative in $f_b\sqrt{\det{g}}$.

Take the spherical harmonic $\ylm(\theta,\phi)$ as the basic deformation,
analagous to the plane wave $\eikr$ in eq.~(\ref{planewave})
or the cylindrical wave $J_m(kr)e^{im\phi}$ in eq.~(\ref{Jmkr}).  The
spherical harmonic is an eigenfunction of $\Delta$ with eigenvalue
$-l(l+1)/R^2$ and an eigenfunction of $\cL$ with eigenvalue
$l(l-1)(l+1)(l+2)/R^4$.  By analogy with the procedure for plane waves
in flat space, we take derivatives as
\begin{eqnarray}
\label{sphere:ULT}
\bu_L=R \nabla \ylm
\hspace{1cm} u_L^\alpha = R D^\alpha \ylm\\
\bu_T=\hat{\br}\times\bu_L
\hspace{1cm} u_T^\alpha=R \epsilon^\alpha_\beta D^\beta \ylm \nonumber
\end{eqnarray}
where $\epsilon$ is the alternating tensor.  We also define
\begin{equation}
\label{sphereUP}
\bu_P=\brh \ylm
\end{equation}
These functions are linear combinations of the ``Vector Spherical
Harmonics'' $\bV_{lm}$, $\bW_{lm}$ and $\bX_{lm}$, which form a
complete set of orthogonal functions for expanding vector fields on
the surface of a sphere~\cite{Hill,Wid86}.  Notice that the transverse mode
$\bu_T$ is proportional to the angular momentum operator acting on
$\ylm$, thus identifying it with the vector spherical harmonic $\bX_{lm}$.
The longitudinal and perpendicular modes, $\bu_L$ and $\bu_P$, are
linear combinations of $\bV_{lm}$ and $\bW_{lm}$.  Note that the
longitudinal and transverse modes $\bu_L$ and $\bu_T$ exist only for
$l\ge 1$, while $\bu_P$ exists for $l\ge 0$.

The transverse mode $\bu_T$ is divergenceless ($u_\gamma^\gamma=0$)
and hence creates no perpendicular force $N$ and no longitudinal force
(the gradient part of $F^\beta$).  In fact, it is an eigenfunction of
the force~(\ref{forces}).  Upon taking into account the commutation of
covariant derivatives, we find $F^\beta=[\mu (l-1)(l+2)/R^2] u_T^\beta$
from which we obtain the eigenvalue
\begin{equation}
\Lambda_T=\mu\frac{(l-1)(l+2)}{R^2}
\end{equation}
As expected, $\Lambda_T=0$ for $l=1$ because these modes correspond to
rigid rotations.

In contrast to the transverse modes, the longitudinal and
perpendicular modes $\bu_L$ and $\bu_P$ are coupled in both the
tangential force $F^\beta$ and perpendicular force $N$.  In matrix
form,
\begin{equation}
\left( \begin{array}{c}F^\alpha \\ N  \end{array} \right) =
\left(\begin{array}{cc} M_{LL} & M_{LP} \\ M_{PL} & M_{PP} \end{array}\right)
\left( \begin{array}{c}u_L^\alpha \\ \zeta  \end{array} \right).
\end{equation}
Setting $\bu_L$ as in eq.~(\ref{sphere:ULT}) and setting $\zeta$ as the
radial component of $\bu_P$ in eq.~(\ref{sphereUP}), the matrix
elements of $M$ become
\begin{equation}
\label{coupling}
\begin{array}{l}
M_{LL}=(\lambda+\mu)\frac{l(l+1)}{R^2}+\mu \frac{(l-1)(l+2)}{R^2}\\ 
M_{LP}=(\lambda+\mu)\frac{2}{R^2} \\
M_{PL}=(\lambda+\mu)\frac{2l(l+1)}{R^2} \\
M_{PP}=(\lambda+\mu)\frac{4}{R^2}+\kappa\frac{(l-1)l(l+1)(l+2)}{R^4}
\end{array}
\end{equation}
The eigenvalues of this matrix, $\lambda_\pm$, are the desired normal
mode eigenvalues $\Lambda$.  For the special case $l=1$ the
eigenvalues are $\lambda_{-}=0$ and $\lambda_{+}=6(\lambda+\mu)/R^2$.
The vanishing eigenvalue $\lambda_{-}$ corresponds to rigid
translation (for example, the north and south pole displace upwards
perpendicular to the shell while the the equator displaces upwards
tangent to the shell).  The finite eigenvalue $\lambda_{+}$
corresponds to an ``optical'' mode in which polar and equatorial
regions displace in opposite directions (for example, the north and
south poles displace upwards while the equator displaces downwards).

The spherical solution should go smoothly to the flat space solution
in polar coordinates as the sphere radius $R\rightarrow\infty$.  This
correspondence can be verified by holding $r=R\theta$, $k=l/R$ and $m$
fixed, and noting~\cite{AS}
\begin{equation}
\lim_{l\rightarrow\infty} \sqrt{\frac{4\pi}{2l+1}} \ylm(\theta,\phi) 
= (-1)^m J_m(k r) e^{im\phi}.
\end{equation}
In addition, the eigenvalues should approach their proper
limits.  Clearly $\Lambda_T$ approaches its flat space
limit~(\ref{flatWT}).  To check $\Lambda_{L,P}$, note that the
eigenvalues $\lambda_{\pm}$ of the matrix~(\ref{coupling}) approach
$(\lambda+2\mu)l(l+1)/R^2$ and $\kappa(l-1)l(l+1)(l+2)/R^4$ in the
limit of large sphere radius $R$, yielding the flat space
limits Eqs.~(\ref{flatWL}) and ~(\ref{flatWP}).

\section{Mass and spring model}
\label{sec:discrete}

We now introduce the discrete mass and spring model for which
numerical calculations will be performed.  This model is also closer
to reality for liposomes and colloidosomes, which consist
respectively, of discrete lipid molecules and colloidal particles, and
also for viruses, which consist of an aggregation of discrete protein
subunits known as capsomers.  Let $\br_i$ be the position of mass $i$
and $\bnh_I$ be the orientation of plaquette $I$.  A plaquette is a
set of three masses joined to each other by springs, and we take the
normal in the outward direction.  Following~\cite{Lidmar03} we define
\begin{equation}
\cH_s=\frac{k_s}{2}\sum_{\langle ij\rangle}(|\br_i-\br_j|-a)^2
\end{equation}
and
\begin{equation}
\cH_b=\frac{k_b}{2}\sum_{\langle IJ\rangle}|\bnh_I-\bnh_J|^2
\end{equation}
and set the unstretched spring length $a=1$.  Here $\langle ij\rangle$
denote pairs of nearest-neighbor vertices, and $\langle IJ\rangle$
denote pairs of adjacent (edge-sharing) plaquettes.  In the continuum
limit the discrete model reproduces the continuum system with elastic
constants
\begin{equation}
\label{eq:Ykappa}
Y=\frac{2}{\sqrt{3}} k_s
\hspace{1cm}
\kappa=\frac{\sqrt{3}}{2} k_b,
\end{equation}
Foppl-von Karman number
\begin{equation}
\label{eq:FvKkskb}
\gamma=\frac{YR^2}{\kappa}=\frac{4k_s R^2}{3k_b},
\end{equation}
Lame coefficients and bulk modulus
\begin{equation}
\label{eq:lammuB}
\lambda=\mu=\frac{\sqrt{3}}{4}k_s
\hspace{1cm} B=\frac{\sqrt{3}}{2}k_s,
\end{equation}
and 2D mass density (taking the vertex mass $m=1$)
\begin{equation}
\rho=2/\sqrt{3}.
\end{equation}

\subsection{Deformations from flat-space}

Consider a regular six-coordinated triangulated network of masses
and springs.  As before we start with the plane wave
function~(\ref{planewave}) and take its gradient to obtain the
longitudinal sound wave. The dispersion relation is simplest for
wavevector $\bk$ in the $\hy$ direction (chosen to lie midway between
two near-neighbor bonds)
\begin{equation}
\Lambda_L=3k_s(1-\cos{(\frac{\sqrt{3}}{2}k_y a)})
\end{equation}
Taking the cross product with $\hz$ yields the transverse sound wave
with dispersion relation
\begin{equation}
\Lambda_T=k_s(1-\cos{(\frac{\sqrt{3}}{2}k_y a)})
\end{equation}
Finally, taking the perpendicular displacements as the planewave
yields the perpendicular modes with dispersion relation
\begin{equation}
\Lambda_P=k_b(2-2\cos{(\frac{\sqrt{3}}{2}k_y a)})^2
\end{equation}
In the continuum limit $ka<<1$ these dispersion relations revert to the
prior results of continuum elastic theory.

\subsection{Buckling of a Plane into a Cone}
\label{sec:cone}

Upon introducing a five-fold $+2\pi/6$ disclination into the flat triangulated
network discussed previously, strain energy accumulates~\cite{Seung88}
and grows without bound as the radius $R$ of the network increases.
At some specific ``buckling radius'' $R_b$ it becomes energetically
favorable to buckle out of plane, trading a reduction in strain energy
for a cost in bending energy.  The trade-off is measured by the
Foppl-von Karman number $\gamma$.  Small $\gamma$ favors flat
networks, while larger $\gamma$ favors buckling into a conical shape.

In the following we analyze the vibrational spectrum of the network as
it passes from flat to conical.  Rather than vary the radius, we hold
$R$ fixed and vary the bending stiffness.  Large $k_b$ opposes
buckling and the network lies flat, while small $k_b$ allows buckling
out of plane into a cone.  For the network of radius $R=8a$ analyzed below,
buckling occurs for $k_b\approx 0.71$.  As $R$ increases the threshold value
of $k_b$ grows as $R^2$ so that $\gamma$ approaches the
limiting value $\gamma_b\approx 154$~\cite{Seung88,Lidmar03}.

\begin{figure*}
\includegraphics*[width=6in]{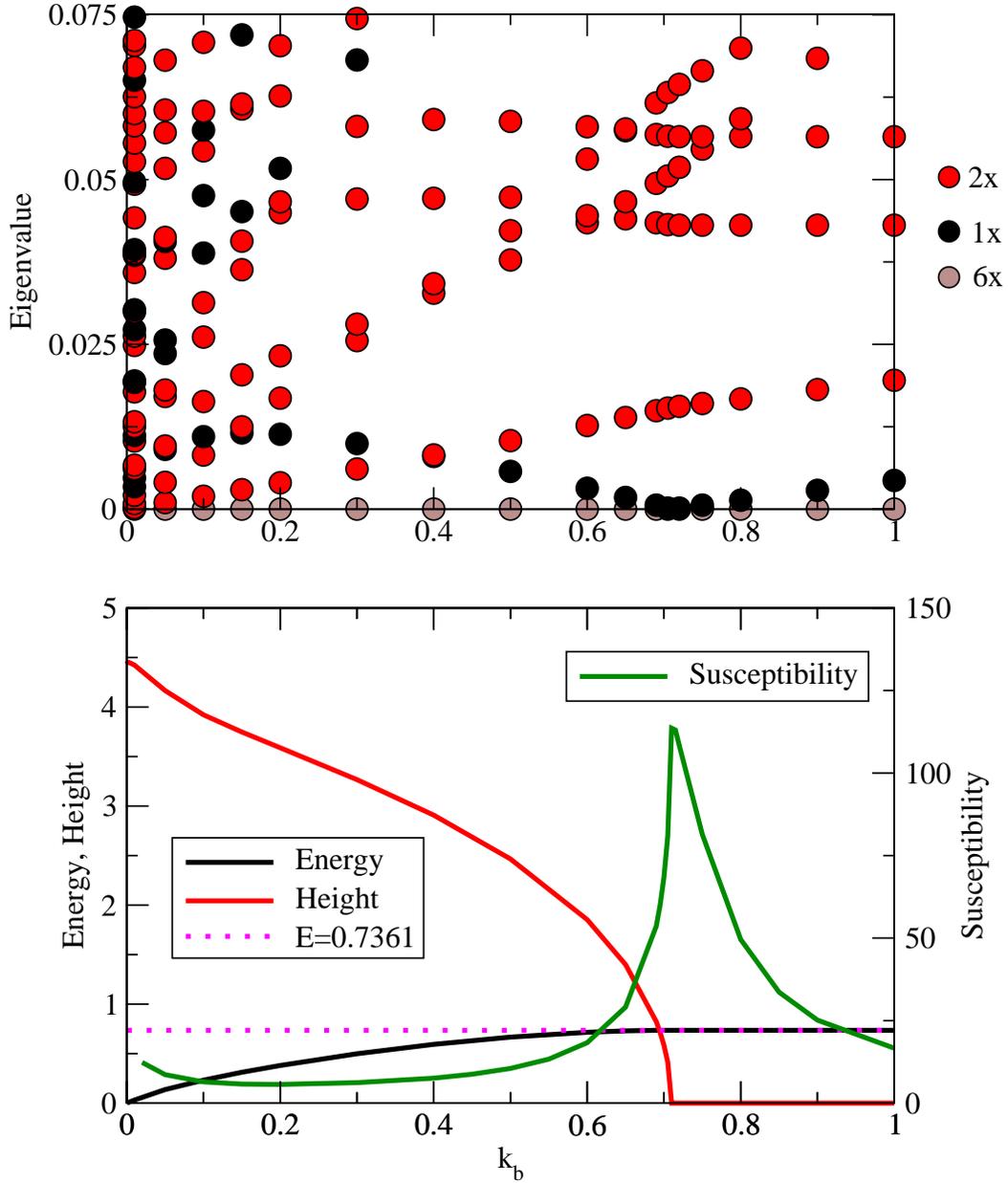}
\caption{Triangulated network of radius $8a$ and spring constant
$k_s=1$ with a single 5-fold disclination at center. (top) Eigenvalue
spectrum color coded according to degeneracy. Note the nondegenerate
1x mode that goes to zero at the buckling transition. (bottom) energy,
cone height and susceptibility.}
\label{fig:spect-cone}
\end{figure*}

Eigenvectors of the Hessian matrix form basis functions for
representations of the symmetry group of the structure~\cite{Tinkham}.
Eigenvectors sharing a common eigenvalue form the basis for an
irreducible representation.  Thus the patterns of degeneracy follow
the dimensionalities of the irreducible representations, as can be
seen in Fig.~\ref{fig:spect-cone}a.  Likewise the eigenvectors exhibit
special symmetry properties associated with subgroups of the full
symmetry group.

The symmetry point group of the cone is $C_{5v}$ in general,
corresponding to five-fold rotations around an axis passing through
the five-coordinated vertex, together with reflections in vertical
planes passing through this axis (see Table~\ref{tab:C5vchar}).  For
the specific case of the flat network, the group is even higher,
$D_{5h}$, adding reflections in the horizontal plane, and two-fold
rotations around axes lying within the plane.  For both groups all
irreducible representations are either 1- or 2-dimensional, so all
nonzero eigenvalues must be nondegenerate or two-fold degenerate.  Of
course, there must be a sixfold degeneracy of zero eigenvalues,
corresponding to rigid translations and continuous rotations (not
belonging to the finite point group) that leave the energy invariant.

\begin{table}
\begin{tabular}{lc|ccccc}
$C_{5v}$ & $m$ & $1C_0$ & $2C_5$ & $2C_5^2$  &5$\sigma_v$ \\
\hline
$A_1$    &   0 &      1 &       1 &       1  &         1  \\
$A_2$    &   0 &      1 &       1 &       1  &        -1  \\
$E_1$    &   1 &      2 &$\tau^{-1}$& $-\tau$&         0  \\
$E_2$    &   2 &      2 & $-\tau$&$\tau^{-1}$&         0  \\
\end{tabular}
\caption{\label{tab:C5vchar} Character table of $C_{5v}$. $C_n$
denotes conjugacy class of order $n$. Values of $m$ denote in-plane
angular momenta.  $\tau=(\sqrt{5}+1)/2$ is the Golden Mean.}
\end{table}

For the group $D_{5h}$, the irreducible representations are based on
those of $C_{5v}$ supplemented with an additional label ${g,u}$
according to whether they are even $(g)$ or odd $(u)$ under reflection
through the horizontal plane $\sigma_h$.  The requirement that each
irreducible representation be either even or odd under $\sigma_h$
requires that each mode be polarized either fully in-plane or fully
perpendicular.

Let $\Lambda_1$ be the lowest nonzero eigenvalue.  Its eigenvector
$\be_1$ is polarized strictly perpendicular to the sheet and
transforms as the irreducible representation $A_{2u}$.  Its value is
nonzero at the origin.  The energy of mode $i$ varies as $\Lambda_i
a_i^2$ where $a_i$ measures the amplitude of the mode.  Mechanical
equilibrium thus demands that all eigenvalues (other than the six zero
modes) be strictly positive.  In particular it requires $\Lambda_1>0$.
However, if we monitor the value of $\Lambda_1$ as a function of
$\gamma$ (Fig.~\ref{fig:spect-cone}a) we find it crosses through zero
at $\gamma_b$.

For small deformations we express the energy as
\begin{equation}
\label{Eeff}
E=\sum_i \frac{1}{2}\Lambda_i a_i^2+{\cal O}(a_i^4)
\end{equation}
Now set $\Lambda_1=c(\gamma_b-\gamma)$.  The mechanically stable
minimum energy structure is perfectly flat ($a_i=0$) for
$\gamma<\gamma_b$, but it buckles out of plane for $\gamma>\gamma_b$,
in a shape described by the eigenvector $\be_1$, with amplitude
growing as $\sqrt{\gamma-\gamma_b}$.  Meanwhile the energy drops as
$(\gamma-\gamma_b)^2$.  These effects can be seen in
figure~\ref{fig:spect-cone}b.

For $\gamma>\gamma_b$, figure~\ref{fig:spect-cone}a shows the spectrum
of vibrations around the mechanically stable, buckled structure.
Note that $\Lambda'_1$ (the lowest nondegenerate eigenvalue) becomes
positive again.

\subsection{Buckling of Spherical Shells}

\subsubsection{$P=1, Q=0$ icosahedron}
\label{sec:ico}

Table~\ref{tab:Ychar} presents the character table of the 60-element
icosahedral rotational symmetry group $Y$, which has 5 irreducible
representations.  The conjugacy classes are labeled $C_n$, where $n$
is the order of elements in the class, so that an element of $C_n$
corresponds to a rotation by $2\pi/n$.  Recall that the spherical
harmonics $Y_{lm}$ form basis functions for the irreducible
representations of the continuous rotation group $SO(3)$, and
therefore they induce representations (in general reducible) of $Y$.
For a given angular momentum $l$ and rotation angle $\theta$, the
character is
\begin{equation}
\chi_l(\theta)=\frac{\sin{(l+1/2)\theta}}{\sin{\theta/2}}.
\end{equation}
Irreducible representations of $Y$ are labeled in
Table~\ref{tab:Ychar} according to the lowest angular momentum $l$
under which they transform.  Of particular interest is the
representation $F_1$ corresponding to angular momentum $l=1$.  This is
the representation under which three-dimensional vectors transform.

\begin{table}
\begin{tabular}{lc|ccccc}
$Y$ & $l$ & $1C_0$ & $15C_2$ & $20C_3$ & $12C_5$ & $12C_5^2$   \\
\hline
$A$ &   0 &      1 &       1 &       1 &       1 &         1   \\
$F_1$&  1 &      3 &      -1 &       0 &  $\tau$ & $-\tau^{-1}$\\
$F_2$& (3)&      3 &      -1 &       0 & $-\tau^{-1}$ & $\tau$ \\
$G$  & (3)&      4 &       0 &       1 &      -1 &        -1   \\
$H$  &  2 &      5 &       1 &      -1 &       0 &         0   \\
\hline
$R$  &    &     12 &       0 &       0 &       2 &         2   \\
$V$  &    &     36 &       0 &       0 & $2\tau$ &$-2\tau^{-1}$\\
\end{tabular}
\caption{\label{tab:Ychar} Character table of $Y$. $C_n$ denotes
conjugacy class of order $n$. Values of $l$ denote angular
momenta. $R$ is the ``regular representation'' and $V$ the ``total
vibrational representation'' discussed in subsection~\ref{sec:ico}}
\end{table}

The simple icosahedron has 12 vertices, 20 faces and 30 edges.
Since we place masses on the vertices, our eigenstates are functions
defined only at vertex positions.  Arbitrary scalar-valued functions
can be expressed as linear combinations of the basis functions of the
``regular representation'', one of which is concentrated at each
icosahedron vertex.  The characters $\chi_R$ of the regular
representation equal the number of vertices that remain stationary
under a given symmetry operation.  Our vibrational modes are {\em
vector}-valued functions on the set of vertices and thus can be
expressed as linear combinations of the product of the regular
representation $R$ times the representation $F_1$ corresponding to a
three dimensional vector.  We call the resulting product
representation the ``total vibrational representation''~\cite{Wid86},
and its characters $\chi_V=\chi_R\chi_{F_1}$.

Reducible representations can be decomposed into their irreducible
components using orthogonality properties of character tables.  In
particular we obtain the decomposition
\begin{equation}
V=A\oplus 3F_1\oplus F_2\oplus 2G\oplus 3H .
\end{equation}
Of the three occurrences of the vector representation $F_1$ we know
that two must correspond to rigid global translations and rotations.
These leave the energy invariant and hence are zero frequency modes.  The
nondegenerate mode transforming as the unit representation $A$ must
correspond to a ``breathing mode'' in which all vertices displace
equally in the radial direction.  We find that the remaining modes
have specific interpretations in terms of vector spherical harmonics,
as listed in table~\ref{tab:modes}.

\begin{table}
\begin{tabular}{lllll}
$\Lambda$ & Formula & Irrep & $g$ &  Comments \\
\hline
0.00000 & 0  & $2\times F_1$ & 6 & Translations + rotations \\
0.58579 & $2-\sqrt{3}$ & $H_a$ & 5 & Mixed contains $\bV_{2m}$ and $\bW_{2m}$\\
0.76393 & $\sqrt{5}/R^2-2$ & $F_2$ & 3 & Radial contains $\br Y_{3m}$ \\
1.00000 & 1  & $H$ & 5 & Tangent $\bX_{2m}$ \\
1.80901 & $1+\tau/2$ & $G_a$ & 4 & Tangent $\bX_{3m}$ \\
2.76393 & $\sqrt{5}/R^2$ & $A$ & 1 & Radial $\br Y_{00}$ breathing mode \\
3.00000 & 3 & $F_1$ & 3 & Mixed $\bV_{1m}$ \\
3.41421 & $2+\sqrt{2}$ & $H_b$ & 5 & Mixed \\
3.42705 & $1+3\tau/2$ & $G_b$ & 4 & Tangent \\
\end{tabular}
\caption{\label{tab:modes} Vibrational eigenvalues for $P=1, Q=0$
icosahedron with $a=k_s=1$, $k_b=0$. $\Lambda$ is eigenvalue and $g$
is degeneracy. $R=\sqrt{1+\tau^2}/2=0.95106$ is the radius of the
icosahedron.}
\end{table}

\subsubsection{Higher Order Icosahedra}

\begin{figure*}[t]
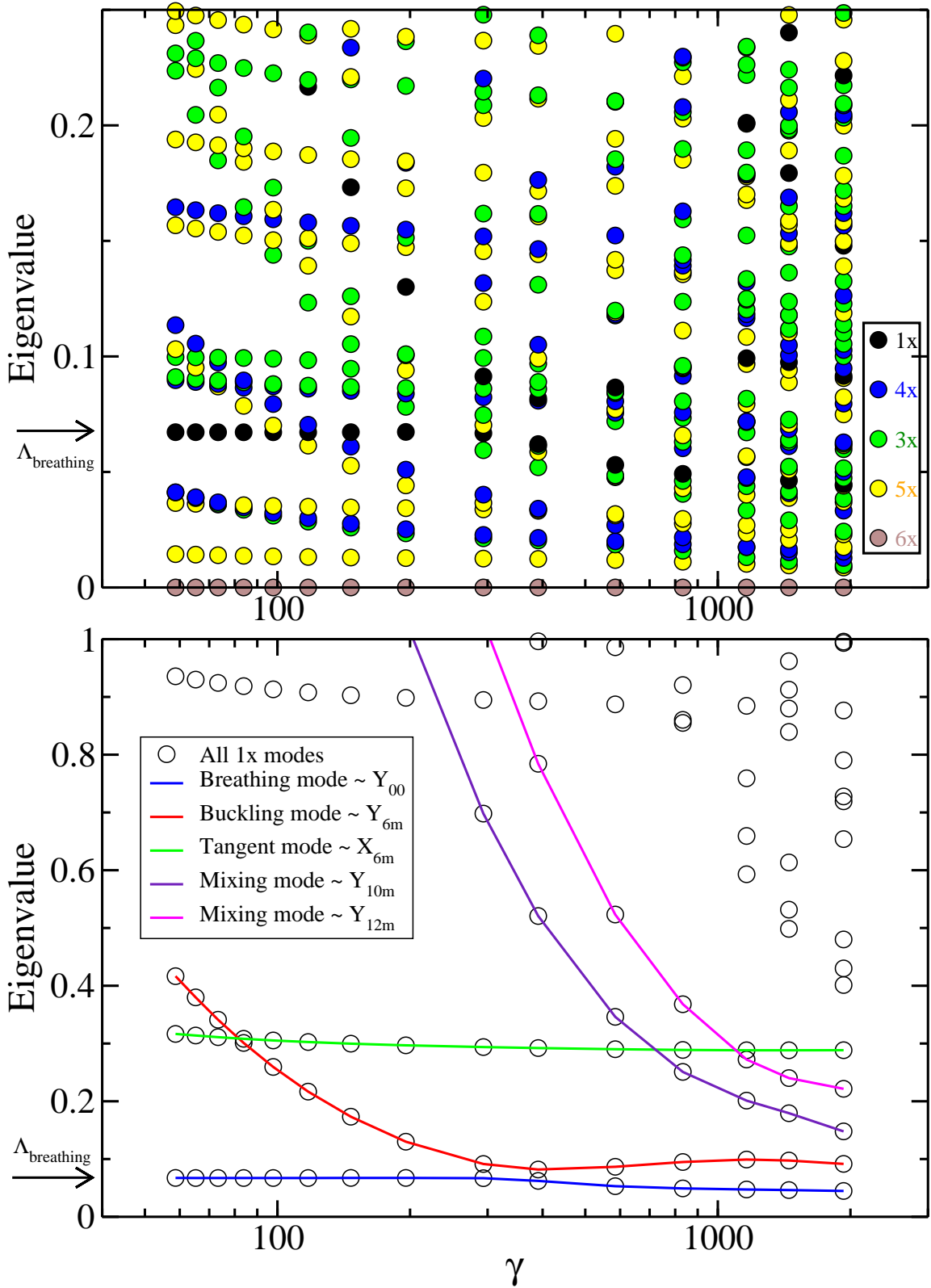

\includegraphics*[width=6in]{spectFvK}
\includegraphics*[width=6in]{1xFvK}
\caption{Lowest frequency modes of $P=8, Q=0$ icosahedron with
$N_v=642$ vertices. (top) Color coded according to
degeneracy. (bottom) Nondegenerate modes only. Arrows locate
eigenvalue $\Lambda_{breathing}=8\pi\sqrt{3}/N_v$. Note the buckling
mode (red) dips close to zero near the buckling transition.}
\label{fig:spect-ico}
\end{figure*}

As the icosahedron is subdivided and the total number of vertices
grows, the classification of modes into irreducible representations
remains similar, but each irreducible representation now occurs many
times.  Fig~\ref{fig:spect-ico}a shows the lowest frequency modes for
a $P=8, Q=0$ icosahedron with $N_v=642$ vertices.  To obtain this
figure, we set $k_s=1$, and varied $k_b$.  For each value of $k_b$ we
relaxed the structure to mechanical equilibrium using steepest
descents, evaluated the Hessian matrix by numerical differentiation,
then diagonalized the matrix.  The Foppl-von Karman number is defined
as in eq.~(\ref{eq:FvKkskb}), where now $R$ is defined as the
root-mean-square radius (defining $R$ instead as the mean
radius~\cite{Lidmar03} has little impact below or near the buckling
transition and results only in a slight rescaling as $\gamma$ grows
large) and takes values in the range 6.6-7.6 for the $P=8, Q=0$
icosahedron.

Owing to rotation and translation invariance of the total energy, we
always have a 6-fold degenerate mode of zero eigenvalue.  The
remaining eigenvalues fall into the classification of icosahedral
symmetry introduced in Table~\ref{tab:Ychar}.

At low $\gamma$, when the shape is spherical in the continuum limit of
large radius, and the energy cost of bending dominates over the energy
cost of stretching or shearing, the lowest frequency nondegenerate
mode is a ``breathing'' mode, corresponding to a sphere with
oscillating radius.  Perturbing the radius by an amount $\zeta$
(i.e. adding mode $\bu_P=\brh \zeta$) increases the energy by $8\pi
B\zeta^2$ while displacing $N_v$ vertices by $\zeta$.  Identifying the
energy with $\frac{1}{2}N_v\Lambda_{breathing}$, and noting the area
modulus $B=\sqrt{3}/2$, we find eigenvalue
$\Lambda_{breathing}=8\pi\sqrt{3}k_s/N_v$ which fits well to the data
in Fig.~\ref{fig:spect-ico}.

\begin{figure}
\includegraphics[width=3in,angle=-90]{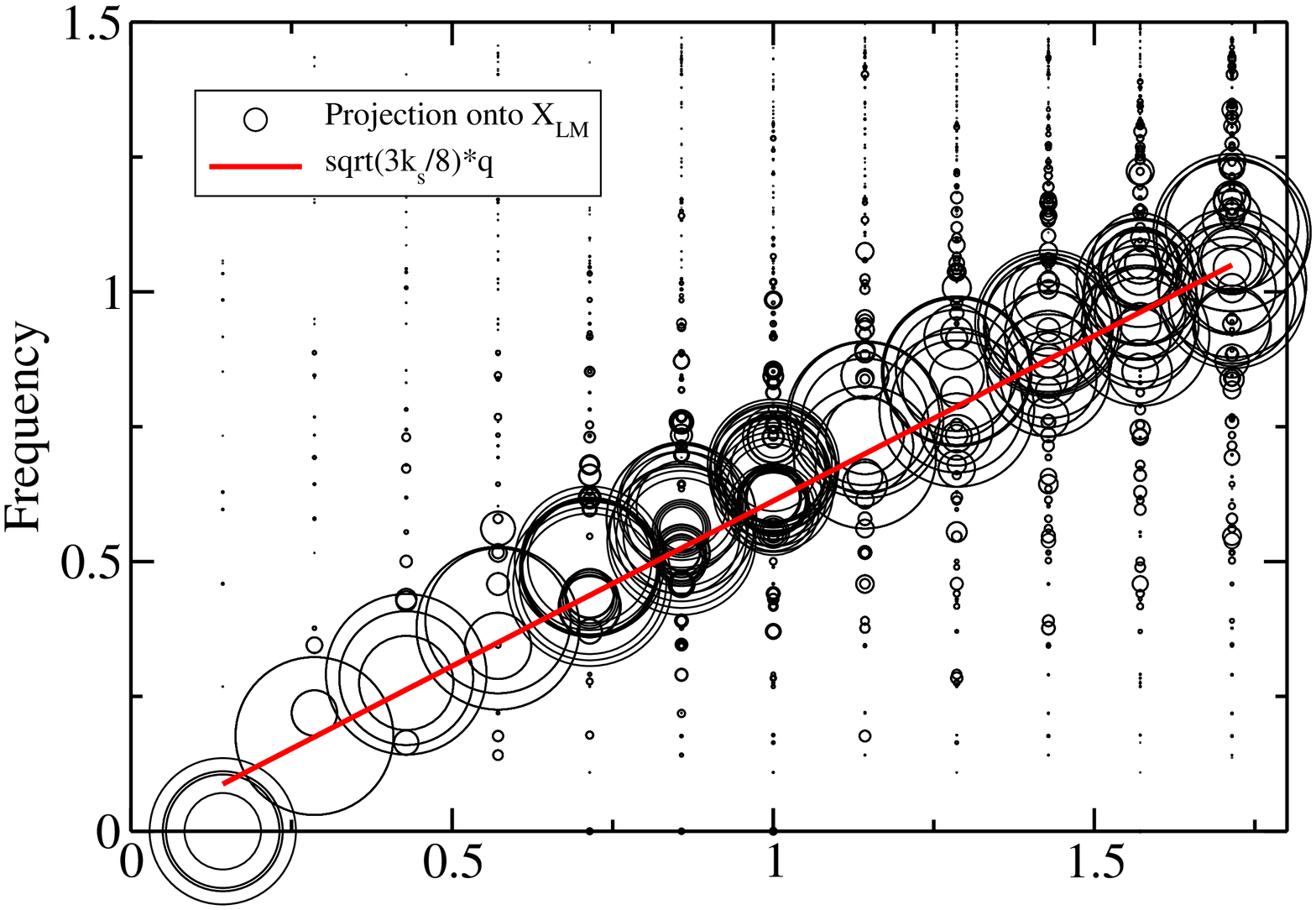}
\includegraphics[width=3in,angle=-90]{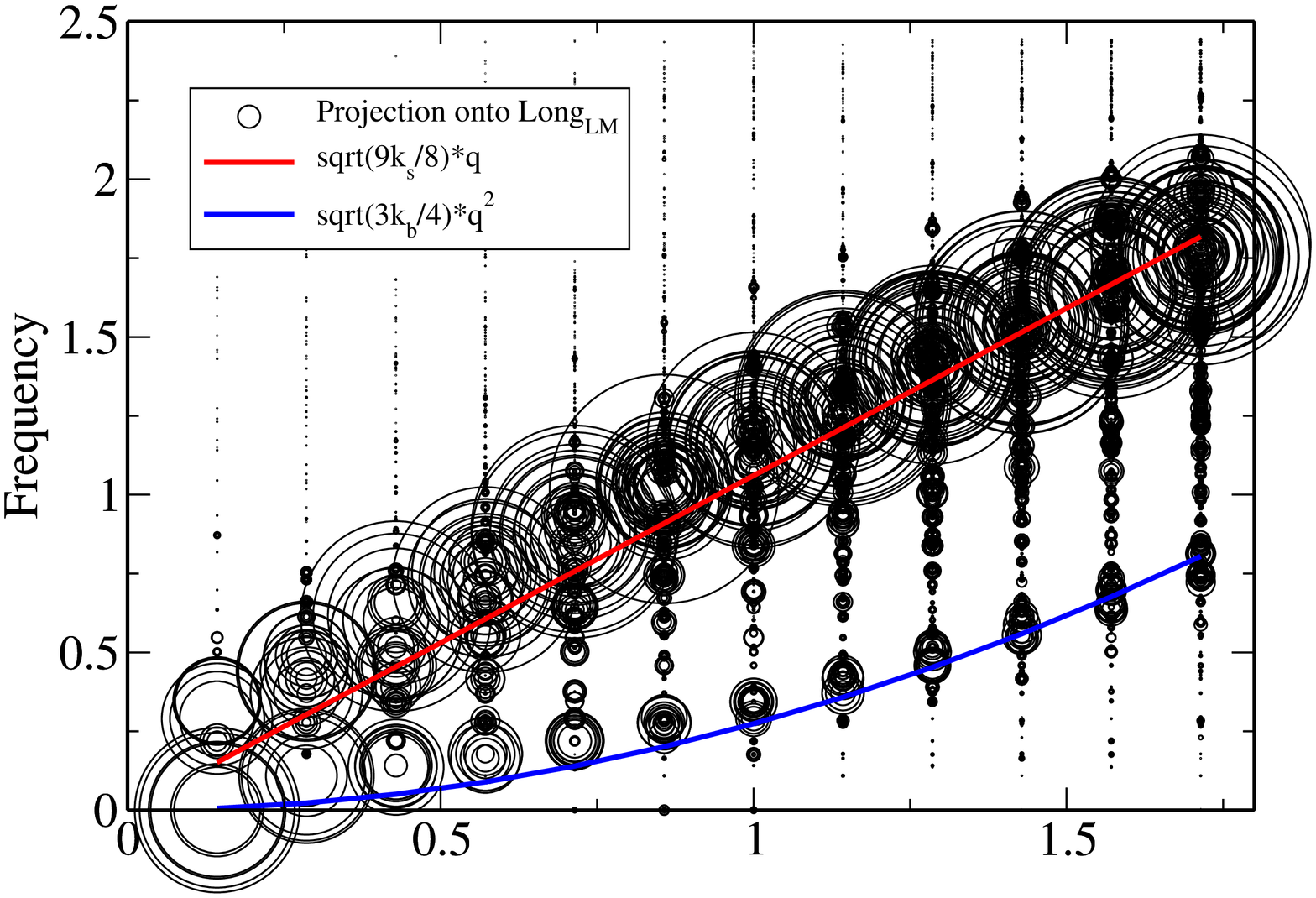}
\includegraphics[width=3in,angle=-90]{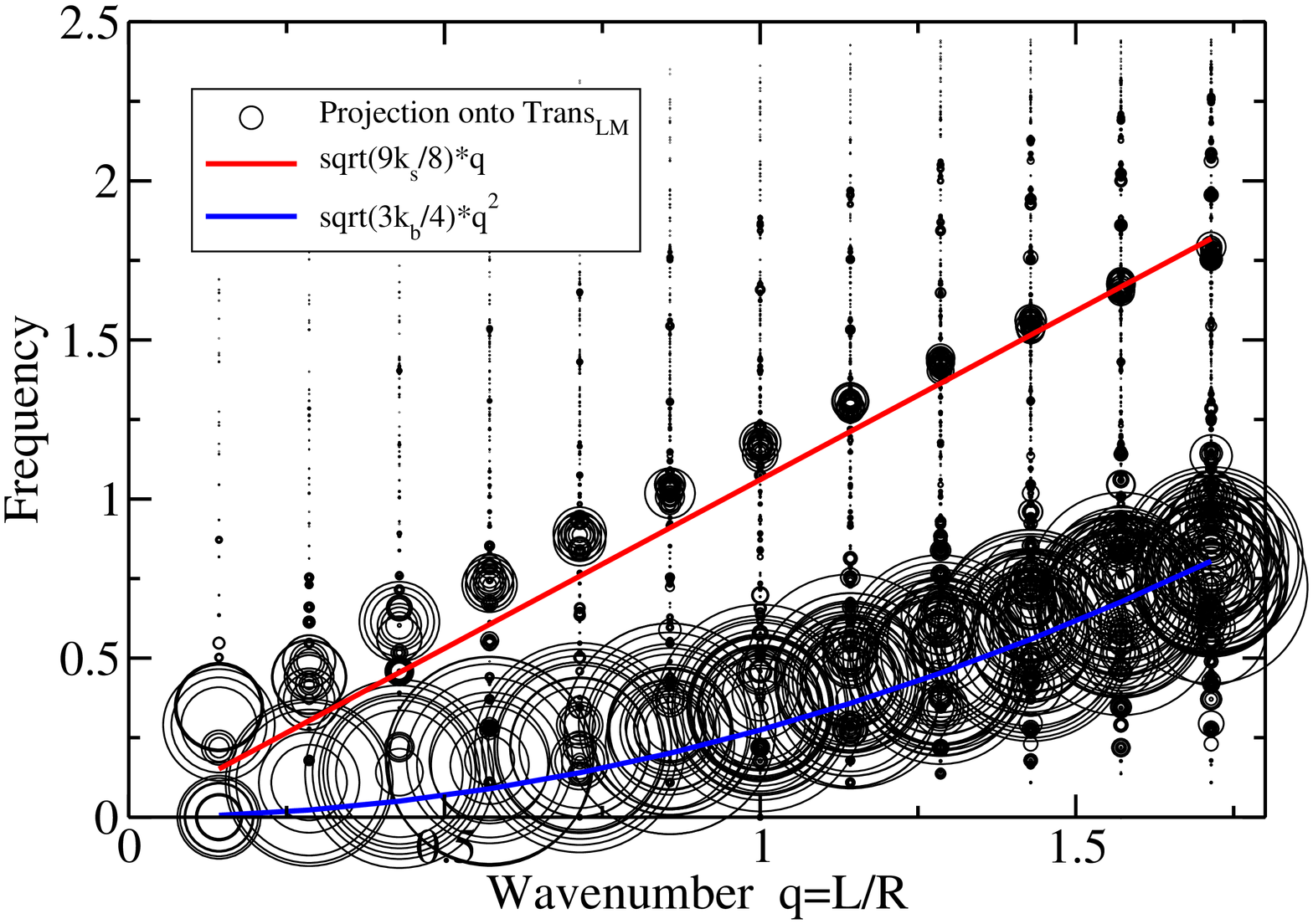}
\caption{\label{fig:disp}Vibrational frequencies plotted versus wave
number $q=l/R$, where $l$ is the angular momentum index. Data is for
$P=8, Q=0$ icosahedron with $k_s=1, k_b=0.1, \gamma=653$.  The radii
of the circles indicate the sizes of the various projections.}
\end{figure}

At higher frequencies, where the wavelength of the modes becomes small
compared to the radius of curvature, we expect that the eigenvalues
should revert to their flat space limits as discussed in
section~\ref{sec:Cont:Sphere}.  The validity of this hypothesis is demonstrated
in the dispersion relations shown in fig.~\ref{fig:disp}.  Here we
plot the vibrational frequencies (i.e. the square roots of Hessian
eigenvalues) as functions of the equivalent wave number, defined as the
angular momentum index $l$ divided by the radius $R$.  The radii of
the circles represent the projections of the eigenvectors onto the
vector spherical harmonics $\bX_{lm}$ (top), and the longitudinal
and transverse eigenfunctions $\bu_L$ and $\bu_T$ (middle and bottom).
The solid lines are the predictions of continuum elastic theory for the
plane, eqs~(\ref{flatWL}-\ref{flatUP}).

Soft-mode behavior at the buckling transition is less pronounced than
in the case of the cone.  The crossover from spherical to faceted
shape, which occurs gradually for $\gamma\sim 100-1000$, preserves the
icosahedral symmetry.  As such, the displacements respect icosahedral
symmetry.  If the transition is due to a ``soft mode'', this mode
itself must be invariant under operations of the icosahedral symmetry
group.  That is, it must transform as the unit representation and
therefore must be nondegenerate.  The soft mode is best seen in
figure~\ref{fig:spect-ico}b, where only the nondegenerate modes are
shown.  Always the lowest frequency nondegenerate mode is an $l=m=0$
``breathing'' mode, and as just discussed its frequency does not
depend significantly on $k_b$.  However, the next occurrence of the
unit representation, at $l=6$, contains a mode, of type $\bu_P$ and
labeled $Y_{6m}$, that does indeed soften and mixes with the breathing
mode in an avoided crossing around $\gamma\approx 400$.  Another $l=6$
mode, of type $\bu_T$ and labeled $\bX_{2m}$, is prevented by symmetry
from mixing with the $\bu_P$ mode.  A series of other nondegenerate
modes $Y_{lm}$ ($l=10, 12, 20, ...$) soften at higher $\gamma$ values
and mix with the other soft modes.

Around $\gamma_b$ the buckling mode consists predominantly of $l=0$
and $l=6$ spherical harmonics, with a small admixture of $l=10$ and
higher harmonics.  The weight of this mode is concentrated in the
vicinities of the icosahedron vertices, and it has strong overlap with
the displacements of vertices under the buckling transition.

The forbidden crossing of the buckling and breathing mode smears of
the buckling transition, because
$\Lambda_{buckling}>\Lambda_{breathing}>0)$ prevents the eigenvalue of
the buckling mode from actually crossing zero.  This contrasts with
the case of the disclinated flat sheet buckling into a cone, where the
eigenvalue does indeed cross zero.  For the sheet-to-cone transition
the analogue of the breathing mode is just a zero energy translation,
rather than a finte frequency radial displacement.  Also, up-down
symmetry of the plane allows the crossing of the buckling mode (which
is odd) with this translation.  On a sphere the symmetry breaking
between inside and outside the sphere causes the breathing mode to mix
with the buckling.

Owing to the smearing, the value of $\gamma_b$ is not uniquely defined
for the sphere-to-icosahedron transition.  Reported values range from
130 to 260 based on fitted energy models~\cite{Lidmar03,Toan06}.
We observe the avoided crossing around $\gamma\approx 400$.

\section{Susceptibilities}
\label{sec:response}

\subsection{Cones}

The soft-mode transition is a genuine sharp phase transition for the
buckling of a disclinated sheet into a cone.  We already discussed the
order parameter (height) and energy variation through the transition,
in section~\ref{sec:cone}.  Now we consider the susceptibility, namely
the response of the order parameter to an applied field. In this case
we examine the response of the buckling height to a point force
applied at the disclination.

Assume the height of the cone (i.e. the vertical displacement of the
5-coordinated particle at the center) is given by $h=\sum_i P_ia_i$,
where again $a_i$ is the amplitude and $P_i$ measures the projection
of the mode $i$ onto the height variable.  Then in the presence of an
applied force we express the energy as
\begin{equation}
\label{Weff}
E=\sum_i(\frac{1}{2}\Lambda_i a_i^2-F P_i a_i),
\end{equation}
which differs from~(\ref{Eeff}) by the work done against the applied force.
Minimizing the energy yields $a_i=FP_i/\Lambda_i$ resulting in total height
$h=F\sum_iP_i^2/\Lambda_i$ 
and susceptibility
\begin{equation}
\chi=\frac{\partial h}{\partial F}=\sum_i \frac{P_i^2}{\Lambda_i}.
\end{equation}
Thus a vanishing eigenvalue, say $\Lambda_1$, passing linearly through
zero at $\gamma=\gamma_b$, translates immediately into a diverging
susceptibility.  This divergence is evident in
Fig.~\ref{fig:spect-cone}b.  Note that the amplitudes differ on the
two sides of $\gamma_b$ because in one case we perturb around a flat
network while in the other we perturb around a buckled cone.

\subsection{Spheres}

\begin{figure*}
\includegraphics*[width=6in,angle=-90]{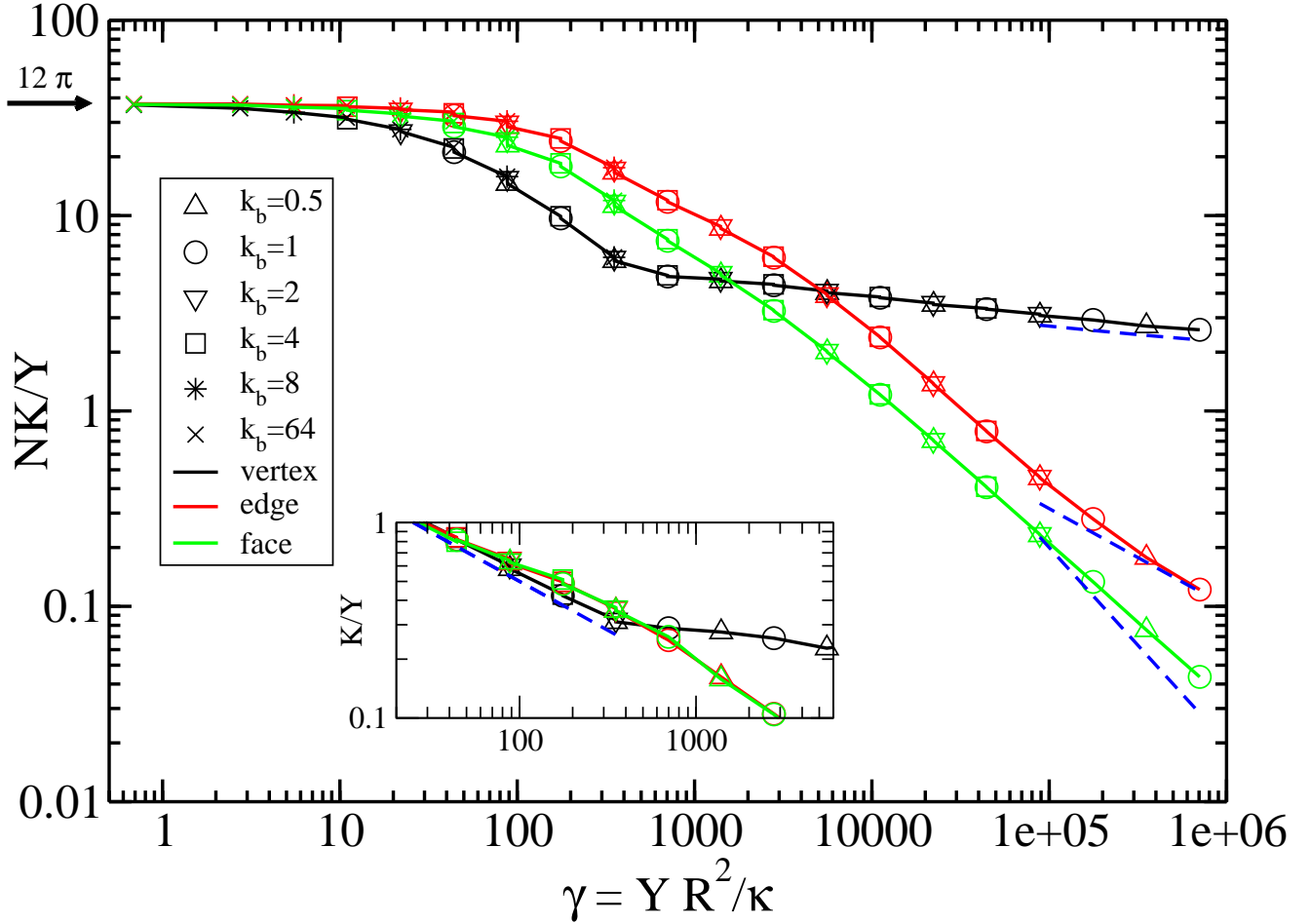}
\caption{Stiffness (inverse susceptibility) to forces applied with
icosahedral symmetry at vertices, edges and faces.  Dashed blue lines
show $15/\log{(0.79\sqrt{\gamma})}$ (vertex, offset from the best fit
for clarity), $100/\sqrt{\gamma}$ (edge, scaling expected for
$\gamma>10^6$), and $20000/\gamma$ (face, scaling for bending of flat
facet~\cite{Timo}).  Inset: diametrically opposed forces applied with uniaxial
symmetry. Dashed blue line shows $5/\sqrt{\gamma}$.}
\label{fig:susc}
\end{figure*}

Now consider the analogous response for the case of icosahedrally
symmetric triangulated spheres and faceted icosahedra.  We consider the
inverse of the susceptibility as an effective spring constant
$K=1/\chi$, and make the spring constant dimensionless by dividing
by the Young's modulus $Y$.  We first present numerical results for
symmetric forces over a wide range of $\gamma$ values, then in later
subsections consider the limiting cases of small and large $\gamma$.

The data was generated from a sequence of icosahedra of varying sizes
and elastic properties.  We consider nonchiral icosahedra with $P=Q$
ranging from 4 to 512 (i.e. $N_v$ ranging from 482 to 7864322).  To
speed calculation, the full 120-element icosahedral symmetry group
$Y_h$ was employed, resulting in a speedup of nearly 120 times.  Beyond
the buckling transition the five-coordinated vertices sharpen, with a
radius of curvature $R_v$ related to the buckling radius
\begin{equation}
\label{eq:Rb}
R_b=\sqrt{\gamma_b\kappa/Y}\approx 12.4\sqrt{\kappa/Y}
\end{equation}
by a geometrical factor of order 1.  In order to approximate the
continuum limit we chose to hold $k_s=1$ fixed and keep $k_b\ge 1/2$,
resulting in $R_b\ge 7.6 a$.

Each structure was relaxed using a conjugate-gradient method.  We
found that the necessary number of relaxation steps diverges with
increasing size, consistent with the $1/R^4$ vanishing relaxation rate
predicted by eq.~(\ref{flatWP}).  To ensure sufficient accuracy in the
susceptibility, we used 128-bit real arithmetic in the final stages of
all relaxations.  For $P=Q=512$, complete relaxation requires
approximately two months on a 3.0GHz Intel Xeon computer.  For studies
such as ours which seek the continuum limit, a finite element
aproach~\cite{DiD01,DiD02,Klug} might be more computationally
efficient than our discrete mass and spring model.

Once the structure was relaxed without applied stress, we re-relaxed
with a radially inward force $F$ applied symmetrically at all $N=12$
vertices, all $N=20$ faces or all $N=30$ edges.  The effective spring
constant $K=F/\zeta$ was defined as the applied force $F$ divided by
the displacement $\zeta$ of the mass to which the force was applied.
We actually consider $NK/Y$ because we define $K$ as the derivative of
$\zeta$ with respect to all $N$ simultaneous applied forces $F$.  Small
applied force $F=0.001$ was required to achieve linear response in
cases where $K$ became small.

Figure~\ref{fig:susc} shows numerical data for symmetric forces
applied to vertices, edges or faces.  In the limit of small $\gamma$
the three data sets converge to a $\gamma$-independent value.  As
$\gamma$ increases, the vertices weaken more quickly than the faces or
edges, consistent with our picture of the buckling transition as
concentrating at the disclinations which are located at vertices.
However, beyond $\gamma_b$ the vertex stiffness falls off very slowly,
while both face and edge stiffness continue their rapid decline.

\subsubsection{Small $\gamma$ limit}

The following discussion first considers the limit of small $\gamma$,
in which the shapes are nearly spherical and calculations can be done
exactly.  The response depends on whether the stress is applied in a
uniaxial manner (e.g. at diametrically opposed points) or in a more
symmetric manner (e.g. applied simultaneously at all vertices or faces
or edges, or even an isotropic pressure).

For an applied pressure $P$ the deformation is purely radial, with
amplitude $\zeta$ as in the breathing mode discussed previously.  This
increases the energy by $8\pi B\zeta^2$, while doing work $4\pi
R^2P\zeta$ against the pressure.  Balancing the two yields
$\zeta=R^2P/4B$, susceptibility $\chi=\partial\zeta/\partial P=R^2/4B$
and spring constant $K/Y=4B/YR^2$.  In the case of $N$ symmetrically
applied point forces, we identify $P=N F/4\pi R^2$ yielding
$NK/Y=16\pi B/Y=12\pi=37.7$, where we used eqs.~(\ref{eq:Ykappa})
and~(\ref{eq:lammuB}).  The stiffness is independent of $\gamma$,
consistent with the numerical result shown.

For uniaxial stress, let the displacement at the two poles be $\zeta$
and assume this displacement persists over a polar region of size $d$
(see section 15 of Ref.~\cite{LandL}).  The bending energy density
$f_b\sim\kappa(\zeta/d^2)^2$, and integrating over the polar region
yields total bending energy $E_b=\kappa\zeta^2/d^2$.  Meanwhile the
strain tensor $u_{\alpha\beta}\sim\zeta/R$ yields a total stretching
energy (see eq.~(\ref{forces})) $E_s\sim Y(\zeta/R)^2d^2$.  Minimizing
the sum $E_s+E_b$ to find the optimal shape yields
$d^4\sim(\kappa/Y)R^2$ and $E_s+E_b\sim\sqrt{\kappa Y}\zeta^2/R$.
Equating this to $F\zeta$, the work done against the applied force, we
find $\zeta\sim(R/\sqrt{\kappa Y})F$ and $\chi=R/\sqrt{\kappa Y}$.
The elastic constant $K/Y\sim 1/\sqrt{\gamma}$, independent of the
axis along which the force is applied, consistent with our numerical
results (see Fig.~\ref{fig:susc}, inset).

\subsubsection{Large $\gamma$ limit}

For $\gamma>\gamma_b$ the radius of curvature at the icosahedron
vertices quickly approaches $R_v$~(eq.~\ref{eq:Rb}) and remains fixed
independent of the icosahedron radius $R$.  Forces applied at
icosahedron vertices get transfered through the curved vertex region
to the flat facets in a primarily longitudinal manner.  According to
the theory of longitudinal deformation of plates (see section 13 of
Ref.~\cite{LandL}) the displacement at large distances $r$ from the
applied force varies as $u(r)\sim (F/Y)\log{r/r_0}$ with $r_0$ some
fixed length.  Upon setting $K=F/u(R)$ and choosing $r_0$ proportional to
$R_v$, we find that $K/Y\sim c/\log{(b\sqrt{\gamma})}$.  The numerical
data shown in Fig.~\ref{fig:susc} fits well to this form with values
$c$=17.3 and $b$=0.79.  The curve shown for comparison illustrates
$c=15$ imposing a uniform displacement for visual clarity.

For forces applied to the icosahedron edges we expect to see the onset
of ridge scaling
behavior~\cite{Wit93,Lob95,DiD01,DiD02,Wood02,Wit07} as
$\gamma$ approaches $10^6$.  Unfortunately the diverging relaxation
time prevents us from exploring larger $\gamma$ within our current
calculational method, preventing us from observing this behavior
cleanly.  We briefly review the predictions of ridge scaling.

Let $L$ (which is proportional to $R$) be the length of an icosahedron
edge.  At each end of this edge the facets join at a fixed angle of
$\theta=138.2^{\circ}$.  At the middle the edge sags inward by an
amount $\zeta$, creating a saddle shaped ridge with a small radius of
curvature $R_1$ across the ridge and a large (and negative) radius of
curvature $R_2$ along the ridge~\cite{Wit93}.  The strain along the
ridgeline is of order $(\zeta/L)^2$.  Because the facets on either
side of the ridge approach angle $\theta$, the radius $R_1$ is
proportional to the sag $\zeta$~\cite{Wit93}.  Assuming that the
bending and strain energy persist along the length $L$ of the ridge
and extend a distance $R_1$ to either side, we estimate the energy as
\begin{equation}
E=R_1 L[Y(\zeta/L)^4+\kappa (1/R_1)^2]-F\zeta
\end{equation}
where the final term represents the action of a force $F$ acting at mid-edge.

Upon setting $\zeta\sim R_1$ and varying $R_1$ to minimize the energy,
we find, in the absence of force $F$,
\begin{equation}
R_1\sim (\kappa/Y)^{1/6}L^{2/3} \sim \sqrt{\kappa/Y}\gamma^{1/3}
\sim L\gamma^{-1/6}.
\end{equation}
In the presence of weak applied force $F$,
the small radius $R_1$ increases by an amount of order
\begin{equation}
\Delta R_1\sim \frac{L}{\sqrt{\kappa Y}} F
\end{equation}
Recalling that $\zeta\sim R_1$ and converting this to an effective
spring constant $K=dF/d\zeta$ yields
$K/Y\sim\sqrt{\kappa/YL^2} \sim 1/\sqrt{\gamma}$.  Indeed, the edge
elasticity in Fig.~\ref{fig:susc} seems to show a crossover towards slope
$-1/2$ on our log-log plot.

Meanwhile, the icosahedron faces become almost planar in
the limit of large $\gamma$.  Timoshenko~\cite{Timo} discusses the
deflection of an equilateral triangular plate under load applied at
the center.  The deflection is proportional to $R^2/\kappa$, from
which we conclude, using eq.~(\ref{eq:FvK}), that $K/Y\sim 1/\gamma$.
However, in Fig.~\ref{fig:susc} the face elasticity seems to follow a
power law closer to -0.8 than -1.  Perhaps residual stresses in the
faces or on their boundaries are responsible for this difference.

\section{Conclusions}

In summary, we investigated the eigenvalue spectrum of a simple
mass-and-spring model of a virus capsid as it passes through its
buckling transition.  The buckling of a spherical shell occurs in a
smooth, nonsingular fashion, in contrast to the buckling of a
disclinated planar network.  The smearing can be attributed to
symmetry-breaking between the interior and exterior of the shell and
is caused by the forbidden crossing of the buckling mode with a lower
frequency breathing mode.

Symmetries of the icosahedron and analogies with continuum elastic
theory were used to classify the normal modes.  Modes of full
icosahedral symmetry, transforming as the unit representation, soften
as the Foppl-von Karman number passes through the buckling transition.
Displacements during buckling, which resemble the maturation of real
virus capsids, can be well represented as a superposition of the two
lowest icosahedrally symmetric modes.  

Susceptibilities to applied forces diverge at the buckling transition
for planar networks.  For spherical topology they evolve smoothly,
with anomalies in the vicinity of $\gamma_b$.  Susceptibility to
forces applied at icosahedron vertices dominates near $\gamma_b$, but
icosahedron edges and faces are much softer for large $\gamma$.  In
the limit of small $\gamma$ the effective spring constant approaches
the behavior of a spherical continuum.

Beyond the buckling transition the faces have the softest linear
response, so this is where one might expect rupture in response to an
isotropic osmotic pressure.  Relative softness of icosahedron faces as
compared to vertices has been reported experimentally in
liposomes~\cite{Delorme}.  We verified this numerically by calculating
the $Q_6$ parameter which measures the distortion from a sphere to an
icosahedron~\cite{Lidmar03}.  Below $\gamma_b$ isotropic pressure
weakly {\em increases} the value of $Q_6$, while above $\gamma_b$
pressure strongly {\em decreases} $Q_6$, bending the facets to make
the shape more nearly spherical.

\begin{acknowledgments}
Work by MW was supported in part by NSF grant DMR-0111198. Work by DRN
was supported by NSF through grant DMR-0231631 and through the Harvard
Materials Research Science and Engineering Center via grant
DMR-0213805.
\end{acknowledgments}

\bibliography{soft}

\end{document}